\documentclass[amsmath,amssymb, twocolumn,aps]{revtex4-2}

\usepackage{graphicx}
\usepackage{dcolumn}
\usepackage{bm}

\begin{document}


\title{Collapse-in and Collapse-out in Partial Measurement in Quantum Mechanics and its WISE Interpretation}

\author{Gui-Lu Long$^{1,2,3,4}$}
 \email{gllong@tsinghua.edu.cn}
 \affiliation{
 $^1$State Key Laboratory of Low-dimensional Quantum Physics and Department of Physics, Tsinghua University, Beijing 100084, China\\
 $^2$Beijing Academy of Quantum Information Sciences, Beijing 100193, China\\
 $^3$Collaborative Innovation Center of Quantum Matter, Beijing 100084, China\\
 $^4$Beijing National Research Center for Information Science and Technology, Beijing 100084, China
 }

\date{\today}
\maketitle

One central issue in quantum mechanics is the relation between the wavefunction and the quantum system it describes. As  quantum mechanics is understood in different ways, the wavefunction is given various explanations. Some regard the wavefunction as "epistemic", that is, something reflected in the human mind, and some regard it as "ontological", i.e., something realistic. The orthodox interpretation of quantum mechanics of Copenhagen \cite{heisenberg1927uber,bohr1928quantenpostulat} is epistemic and treats the wavefunction merely as a mathematical quantity. Recent examples of the ontological interpretation include the random discontinuous motion \cite{gao1999interpretation},  "Wavefunction Is the System Entity"(WISE) interpretation \cite{gui2006general}, and information complete interpretation \cite{chen2014information}. WISE treats the wavefunction equivalently as the quantum system itself, that is, the quantum system is just the wavefunction, and the wavefunction is just the quantum system. These two are exactly the same. The quantum system, which is also the wavefunction, can exist in disjoint regions of space, travel at a finite speed, and collapse upon measurements. An encounter-delayed-choice experiment has been proposed and experimentally demonstrated recently \cite{long2018realistic}.

In this short communication, we will concentrate on the partial measurement issue and give an explanation concerning the WISE interpretation. The essential idea of WISE is given in Ref. \cite{gui2006general}, together with the linear combination of unitaries (LCU) formalism of quantum computing. LCU has now become one of the major techniques in quantum algorithm design. The quantum circuit implementation of LCU  is given in Refs. \cite{gui2008duality,gui2009allowable}, and  a review of the subject is given in Ref.\cite{long2011duality}.

{ \bf Partial measurement postulate.} We recall first the measurement postulate in standard quantum mechanics.  If a particle is in state $|\Psi\rangle$, a measurement of the variable (corresponding to) $\Omega$ will yield one of the eigenvalues $\omega$ with probability $P(\omega)\propto |\langle \omega|\Psi\rangle|^2$. The state of the system will change from $|\Psi\rangle$ to $|\omega\rangle$ as a result of the measurement \cite{ramamurti1994principles}.

What will happen if the measurement is on part of the wave function (partial measurement) rather than on a full wave function (full measurement)? For instance, in a three-slits system, if one places a detector immediately after the first slit and places no detectors in the remaining two slits,  then a partial measurement is established. Here we give the details of the partial measurement.

Using the quantum circuit realization in Refs. \cite{gui2008duality,gui2009allowable}, the wavefunction of an electron passing through a $d$-slit is represented as $|\Phi\rangle=\sum_{i=1}^d c_i|\psi_i\rangle|i\rangle$, where $|i\rangle$ represents the $i$-th slit, and $|\psi_i\rangle $ is the sub-wavefunction in the $i$-th slit. For simplicity, we assume that $|\psi_i\rangle=|\psi\rangle$. The wavefunction is normalized, namely $\sum_i |c_i|^2=1$.

  In the WISE interpretation, these sub-waves as a whole form an "electron". Thus, it is easy to comprehend that an "electron"
passes through the $d$-slits simultaneously. The "electron" is no longer a rigid sphere. It is distributed in space, even disjointedly. It changes its shape as the wavefunction changes.

If one places  a  detector just after slit-1, then there is a probability $|c_1|^2$ that the detector will measure the electron, and the whole wavefunction will collapse into slit-1. What would happen if the detector at slit-1 does not get a result? To this end,  we rewrite $|\Psi\rangle$ as
\begin{eqnarray}
|\Phi\rangle=c_1|\psi\rangle|1\rangle + \sqrt{1-|c_1|^2}|\psi\rangle |S2\rangle,\label{e2}
\end{eqnarray}
where
\begin{eqnarray}
|S2\rangle= \sum_{i=2}^d {c_i\over \sqrt{1-|c_1|^2}} |i\rangle. \label{e2s}
\end{eqnarray}
Eq. (\ref{e2}) is like a double-slit experiment with slit-1 and slit-S2. If we imaginarily place two detectors, one at slit-1 and one at slit-S2, then there is a probability $|c_1|^2$ that the detector will measure the electron at slit-1, and the whole wavefunction will collapse into slit-1, and a probability 1-$|c_1|^2$  that the detector at nominal slit-S2 will measure the electron. By this heuristic reasoning, we give our partial measurement postulate, which is a novel development of the measurement postulate of standard quantum mechanics:

 Suppose a quantum system is in state $\sum_{i=1}^M c_i|\omega_i\rangle +\sum_{j=M+1}^d u_j |\omega_j\rangle$, where $\sum_i{ |c_i|^2}+\sum_{j}{|u_j|^2}=1$
,  $\omega_i$
is one of the eigenvalues of observable $\Omega$, and $|\omega_i\rangle$ is the corresponding eigenvector. If part of the wavefunction, $\sum_{i=1}^M c_i|\omega_i\rangle$,
is measured in variable $\Omega$ , then the result of the measurement will be one of the following:

 (1) Collapse-in: One eigenvalue  $\omega_i$ will be obtained  with probability $|\langle \omega_i|\Psi\rangle|^2=|c_i|^2$, where $1\le i \le M$. After the measurement, the state of the system will change instantly from $|\Psi\rangle$ into $|\omega_i\rangle$.

 (2) Collapse-out: The part of the wavefunction being measured will disappear, and the state of the system will change instantly to the unmeasured part, namely, with probability $1-\sum_{i=1}^M|c_i|^2$,
 \begin{eqnarray}
 |\Psi\rangle\rightarrow {1\over \sqrt{1-\sum_{i=1}^M|c_i|^2}}\sum_{j=M+1}^d u_j |\omega_j\rangle.
 \end{eqnarray}

 Collapse-in and collapse-out of partial measurement happen randomly not only in space, but also over time. Though it is seldom discussed, partial measurement appears very often in reality.  For instance,  the detection of a photon by a detector can be naturally understood in terms of this partial measurement postulate. When the wavefunction of a photon goes to a detector, it is not measured as a whole at the same time. Its front part arrives at the detector first, hitting some area of the detector. It can collapse-in at any point of the intersecting area, with respective probabilities (randomly in space). If the collapse-in does not happen, then collapse-out happens instead. The front part of the wavefunction will disappear, and the corresponding probability will be shifted to another part of the wavefunction. At the next instant, collapse-in or collapse-out happens again.  This process continues until the photon is detected. If the photon has not been detected until the last part of the wavefunction reaches the detector, then the amplitude of this remaining wavefunction increases to the full so as to give a probability 1, so that the photon will be surely detected at the final step.

 {\it \bf Measurement is reaction.} The essential process of a measurement is an enhanced and concentrated reaction.  A photon may react with a single atom in free space with a tiny probability. In a detector, a huge number of such atoms are concentrated in a small area, thus the probability of such reaction is increased tremendously. Of course, in addition to such concentration effect, there exist other effects in a detecting process, such as the avalanche effect in a photon detector.

 When reacting with other quantum systems, a quantum system takes part in as a whole according to the measurement postulate, rather than just a part of the quantum system.  Schr\"{o}dinger once wanted to treat the wavefunction of an electron as an electron cloud, but finally abandoned it because no partial electric charge could be found.  If we make an assumption that a quantum system takes part in a reaction as a whole, then this difficulty can be overcome easily. The WISE interpretation goes further to treat the wavefunction of a quantum system as the quantum system itself, thus answering similar questions like why no fractional electric charge is found. This wholeness nature of a quantum system is also important in understanding why a photon can preserve its properties after being generated for many years and traveling a long distance of many light years: if it reacts with another quantum system in an interstellar matter on its way to the Earth, it will react wholly (leading to a collapse-in measurement); otherwise, it will retain its properties (collapse-out, with part of its wavefunction "bitten" by the encountered interstellar matter, and with respective probabilities shifted to other parts of the wavefunction) and continue its way to the Earth.

 The measurement process takes place randomly in space over time. It is NOT governed by Schr\"{o}dinger equation. Actually, it is governed by something beyond quantum mechanics. Thus, in the view of the WISE interpretation, the world is a mixture of determinism and randomness. All quantum systems are evolving according to  Schr\"{o}dinger equations with respective interactions. Because of the inter-interactions between different systems, reactions occur with varied probabilities, randomly in space over time. These reactions are objective.

This work was supported by the National Key Research and Development Program of China under Grant No.2017YFA0303700, National Natural Science Foundation of China under Grant No.11974205, and  Beijing Advanced Innovation Center for Future Chip (ICFC) and Tsinghua University Initiative Scientific Research Program.

\bibliography{citref}

\providecommand{\noopsort}[1]{}\providecommand{\singleletter}[1]{#1}%
\begin{thebibliography}{10}%
\makeatletter
\providecommand \@ifxundefined [1]{%
 \@ifx{#1\undefined}
}%
\providecommand \@ifnum [1]{%
 \ifnum #1\expandafter \@firstoftwo
 \else \expandafter \@secondoftwo
 \fi
}%
\providecommand \@ifx [1]{%
 \ifx #1\expandafter \@firstoftwo
 \else \expandafter \@secondoftwo
 \fi
}%
\providecommand \natexlab [1]{#1}%
\providecommand \enquote  [1]{``#1''}%
\providecommand \bibnamefont  [1]{#1}%
\providecommand \bibfnamefont [1]{#1}%
\providecommand \citenamefont [1]{#1}%
\providecommand \href@noop [0]{\@secondoftwo}%
\providecommand \href [0]{\begingroup \@sanitize@url \@href}%
\providecommand \@href[1]{\@@startlink{#1}\@@href}%
\providecommand \@@href[1]{\endgroup#1\@@endlink}%
\providecommand \@sanitize@url [0]{\catcode `\\12\catcode `\$12\catcode
  `\&12\catcode `\#12\catcode `\^12\catcode `\_12\catcode `\%12\relax}%
\providecommand \@@startlink[1]{}%
\providecommand \@@endlink[0]{}%
\providecommand \url  [0]{\begingroup\@sanitize@url \@url }%
\providecommand \@url [1]{\endgroup\@href {#1}{\urlprefix }}%
\providecommand \urlprefix  [0]{URL }%
\providecommand \Eprint [0]{\href }%
\providecommand \doibase [0]{https://doi.org/}%
\providecommand \selectlanguage [0]{\@gobble}%
\providecommand \bibinfo  [0]{\@secondoftwo}%
\providecommand \bibfield  [0]{\@secondoftwo}%
\providecommand \translation [1]{[#1]}%
\providecommand \BibitemOpen [0]{}%
\providecommand \bibitemStop [0]{}%
\providecommand \bibitemNoStop [0]{.\EOS\space}%
\providecommand \EOS [0]{\spacefactor3000\relax}%
\providecommand \BibitemShut  [1]{\csname bibitem#1\endcsname}%
\let\auto@bib@innerbib\@empty
\bibitem [{\citenamefont {Heisenberg}(1927)}]{heisenberg1927uber}%
  \BibitemOpen
  \bibfield  {author} {\bibinfo {author} {\bibfnamefont {W.}~\bibnamefont
  {Heisenberg}},\ }\bibfield  {title} {\bibinfo {title} {Uber den anschaulichen
  inhalt der quanten theoretischen kinematik und mechanik},\ }\href@noop {}
  {\bibfield  {journal} {\bibinfo  {journal} {Zeit. f. Phyzik}\ }\textbf
  {\bibinfo {volume} {43}},\ \bibinfo {pages} {172} (\bibinfo {year}
  {1927})}\BibitemShut {NoStop}%
\bibitem [{\citenamefont {Bohr}(1928)}]{bohr1928quantenpostulat}%
  \BibitemOpen
  \bibfield  {author} {\bibinfo {author} {\bibfnamefont {N.}~\bibnamefont
  {Bohr}},\ }\bibfield  {title} {\bibinfo {title} {Das quantenpostulat und die
  neuere entwicklung der atomistik},\ }\href@noop {} {\bibfield  {journal}
  {\bibinfo  {journal} {Naturwissenschaften}\ }\textbf {\bibinfo {volume}
  {16}},\ \bibinfo {pages} {245} (\bibinfo {year} {1928})}\BibitemShut
  {NoStop}%
\bibitem [{\citenamefont {Gao}(1999)}]{gao1999interpretation}%
  \BibitemOpen
  \bibfield  {author} {\bibinfo {author} {\bibfnamefont {S.}~\bibnamefont
  {Gao}},\ }\bibfield  {title} {\bibinfo {title} {The interpretation of quantum
  mechanics (i) and (ii)},\ }\href@noop {} {\bibfield  {journal} {\bibinfo
  {journal} {arXiv preprint physics/9907001}\ } (\bibinfo {year}
  {1999})}\BibitemShut {NoStop}%
\bibitem [{\citenamefont {Long}(2006)}]{gui2006general}%
  \BibitemOpen
  \bibfield  {author} {\bibinfo {author} {\bibfnamefont {G.-L.}\ \bibnamefont
  {Long}},\ }\bibfield  {title} {\bibinfo {title} {General quantum interference
  principle and duality computer},\ }\href@noop {} {\bibfield  {journal}
  {\bibinfo  {journal} {Communications in Theoretical Physics}\ }\textbf
  {\bibinfo {volume} {45}},\ \bibinfo {pages} {825} (\bibinfo {year}
  {2006})}\BibitemShut {NoStop}%
\bibitem [{\citenamefont {Chen}(2014)}]{chen2014information}%
  \BibitemOpen
  \bibfield  {author} {\bibinfo {author} {\bibfnamefont {Z.-B.}\ \bibnamefont
  {Chen}},\ }\bibfield  {title} {\bibinfo {title} {The information-complete
  quantum theory},\ }\href@noop {} {\bibfield  {journal} {\bibinfo  {journal}
  {arXiv preprint arXiv:1412.1079}\ } (\bibinfo {year} {2014})}\BibitemShut
  {NoStop}%
\bibitem [{\citenamefont {Long}\ \emph {et~al.}(2018)\citenamefont {Long},
  \citenamefont {Qin}, \citenamefont {Yang},\ and\ \citenamefont
  {Li}}]{long2018realistic}%
  \BibitemOpen
  \bibfield  {author} {\bibinfo {author} {\bibfnamefont {G.}~\bibnamefont
  {Long}}, \bibinfo {author} {\bibfnamefont {W.}~\bibnamefont {Qin}}, \bibinfo
  {author} {\bibfnamefont {Z.}~\bibnamefont {Yang}},\ and\ \bibinfo {author}
  {\bibfnamefont {J.-L.}\ \bibnamefont {Li}},\ }\bibfield  {title} {\bibinfo
  {title} {Realistic interpretation of quantum mechanics and
  encounter-delayed-choice experiment},\ }\href@noop {} {\bibfield  {journal}
  {\bibinfo  {journal} {SCIENCE CHINA Physics, Mechanics \& Astronomy}\
  }\textbf {\bibinfo {volume} {61}},\ \bibinfo {pages} {030311} (\bibinfo
  {year} {2018})}\BibitemShut {NoStop}%
\bibitem [{\citenamefont {Long}\ and\ \citenamefont
  {Liu}(2008)}]{gui2008duality}%
  \BibitemOpen
  \bibfield  {author} {\bibinfo {author} {\bibfnamefont {G.-L.}\ \bibnamefont
  {Long}}\ and\ \bibinfo {author} {\bibfnamefont {Y.}~\bibnamefont {Liu}},\
  }\bibfield  {title} {\bibinfo {title} {Duality computing in quantum
  computers},\ }\href@noop {} {\bibfield  {journal} {\bibinfo  {journal}
  {Communications in Theoretical Physics}\ }\textbf {\bibinfo {volume} {50}},\
  \bibinfo {pages} {1303} (\bibinfo {year} {2008})}\BibitemShut {NoStop}%
\bibitem [{\citenamefont {Long}\ \emph {et~al.}(2009)\citenamefont {Long},
  \citenamefont {Liu},\ and\ \citenamefont {Wang}}]{gui2009allowable}%
  \BibitemOpen
  \bibfield  {author} {\bibinfo {author} {\bibfnamefont {G.-L.}\ \bibnamefont
  {Long}}, \bibinfo {author} {\bibfnamefont {Y.}~\bibnamefont {Liu}},\ and\
  \bibinfo {author} {\bibfnamefont {C.}~\bibnamefont {Wang}},\ }\bibfield
  {title} {\bibinfo {title} {Allowable generalized quantum gates},\ }\href@noop
  {} {\bibfield  {journal} {\bibinfo  {journal} {Communications in Theoretical
  Physics}\ }\textbf {\bibinfo {volume} {51}},\ \bibinfo {pages} {65} (\bibinfo
  {year} {2009})}\BibitemShut {NoStop}%
\bibitem [{\citenamefont {Long}(2011)}]{long2011duality}%
  \BibitemOpen
  \bibfield  {author} {\bibinfo {author} {\bibfnamefont {G.-L.}\ \bibnamefont
  {Long}},\ }\bibfield  {title} {\bibinfo {title} {Duality quantum computing
  and duality quantum information processing},\ }\href@noop {} {\bibfield
  {journal} {\bibinfo  {journal} {International Journal of Theoretical
  Physics}\ }\textbf {\bibinfo {volume} {50}},\ \bibinfo {pages} {1305}
  (\bibinfo {year} {2011})}\BibitemShut {NoStop}%
\bibitem [{\citenamefont {Shankar}(1994)}]{ramamurti1994principles}%
  \BibitemOpen
  \bibfield  {author} {\bibinfo {author} {\bibfnamefont {R.}~\bibnamefont
  {Shankar}},\ }\href@noop {} {\emph {\bibinfo {title} {Principles of Quantum
  Mechanics, 2nd edition}}}\ (\bibinfo  {publisher} {Plenum Press New York},\
  \bibinfo {year} {1994})\BibitemShut {NoStop}%
\end{thebibliography}%

\end{document}